\documentclass[conference,a4paper]{IEEEtran}

\usepackage{xcolor}
\usepackage{balance}
\usepackage{booktabs}
\usepackage{cite}
\usepackage{amsmath,amssymb,amsfonts}
\usepackage{algorithmic}
\usepackage{graphicx}
\usepackage{textcomp}
\usepackage{array}
\usepackage{multirow}
\usepackage{balance}

\usepackage{pgfplots}
\usepackage{tikz}
\usetikzlibrary{3d}
\pgfplotsset{compat=newest}

\usepackage{steinmetz}

\usepackage{cite}
\usepackage{amsmath,amssymb,amsfonts}
\usepackage{algorithmic}
\usepackage{graphicx}
\usepackage{textcomp}
\usepackage{caption}
\usepackage{subcaption}
\usepackage{steinmetz}
\usepackage{color}
\usepackage{circuitikz}
\usepackage{xcolor}
\usepackage{array}
\tikzset{>=latex} 
\definecolor{myblue}{rgb}{0, 0.4470, 0.7410}
\definecolor{mygray}{rgb}{0.7, 0.7, 0.7}
\usepackage{balance}
\usepackage{breqn}

\ifCLASSOPTIONcompsoc
 \usepackage[caption=false,font=normalsize,labelfont=sf,textfont=sf]{subfig}
\else
 \usepackage[caption=false,font=footnotesize]{subfig}
\fi

\begin{document}
\title{Compact Meandered Dipole Array Presenting~Super-Realized Gain}

\author{\IEEEauthorblockN{
Donal Patrick Lynch\IEEEauthorrefmark{1},   
Mengran Zhao\IEEEauthorrefmark{1}, 
Aaron M. Graham\IEEEauthorrefmark{1},
Stylianos D. Asimonis\IEEEauthorrefmark{2}
}

\IEEEauthorblockA{\IEEEauthorrefmark{1}
Centre for Wireless Innovation (CWI), Queens University Belfast, Queen’s Road, Belfast BT3 9DT, UK.}
\IEEEauthorblockA{\IEEEauthorrefmark{2}
Department of Electrical and Computer Engineering,
University of Patras, GR26500 Patras, Greece.}
\IEEEauthorblockA{ \emph{dlynch27@qub.ac.uk, mengran.zhao@qub.ac.uk, s.asimonis@upatras.gr}
}}

\maketitle

\begin{abstract}
This paper presents a comprehensive study of a compact two-element meandered dipole array aimed at achieving super-realized gain. An optimization was performed using a Genetic Algorithm (GA) implemented in MATLAB, targeting the maximization of the realized gain in the end-fire direction. Three primary excitation schemes were evaluated: the conventional optimization to achieve superdirectivity (signal magnitude and phase angle), phase-only optimization, and uniform excitation (characterized by equal amplitude and no phase shift). The results show that by carefully optimizing both the signal magnitudes and phase angles, the array could achieve a substantial improvement in realized gain. Phase-only optimization provided a competitive realized gain with only minor reductions compared to conventional optimization, suggesting that optimizing the signal phase alone can be an effective strategy in practical implementations.
\end{abstract}

\vskip0.5\baselineskip

\section{Introduction}
The rapid development of wireless communication technologies has intensified the need for compact high performance antennas that can be integrated into small devices without sacrificing functionality. Applications such as IoT (Internet of Things), wearable devices, and wireless sensor networks require antenna arrays that not only occupy minimal physical space but also present high directivity and high radiation efficiency, while maintaining good impedance matching. Designing antennas that meet these requirements presents a significant challenge, particularly when dealing with compact antennas, which typically suffer from reduced efficiency and narrow bandwidth \cite{wong2002compact}.

Meandered dipoles have emerged as a promising solution to the challenges of compact antenna design \cite{Marrocco2003, song2024}. These designs reduce the physical length of an antenna while preserving its electrical length, enabling compact configurations without significantly compromising performance. Uskov showed that a uniform linear array (ULA) of $N$ isotropic radiators can achieve end-fire directivity increases to $N^2$ as the spacing between elements approaches zero\cite{uzkov}. Hansen followed this and defined the concept of 'Superdirectivity': an array is considered superdirective if its directivity surpasses that of an identical array with uniform excitation \cite{Hansen1981Fundamental}.

Achieving superdirectivity in practice, however, requires careful management of factors such as element spacing, mutual coupling, and precise element excitation \cite{debard2023foldeddipole,haskou2015printedhalfloop,lynch2023sdhuygens,moore2024aaron, graham2024rectenna}. Physical realizations of superdirective arrays, such as those demonstrated by Altshuler \cite{Altshuler2005monopole}, have proven its feasibility, though they rely on attenuators and amplifiers to finely adjust the magnitudes and phases of the monopole elements. An alternative approach to improve realized gain involves slightly modifying the dimensions of half-wavelength dipoles and exciting each with equal amplitude and optimal phase \cite{Lynch2024}. Another approach is shown \cite{Assimonis2023How} which employs a loaded dipole configuration, and focuses on the optimal load values, showing potential for practical superdirective antennas. Clemente presents a four-element superdirective meandered dipole array that achieved a directivity of $10$~dBi and a realized gain of $0.84$~dBi, demonstrating that it is possible to attain superdirectivity even with a compact design \cite{clemente2017meander}. 

The objective of this paper is to achieve a compact dipole array that presents super-realized gain by reducing the physical length of the elements through meandering, while maintaining high directivity, high radiation efficiency, and impedance matching to $50$~$\Omega$. This paper is structured as follows: Section~II provides an overview of the optimization setup, detailing the Genetic Algorithm employed for maximizing the realized gain of the meandered dipole array. Section~III presents the results from the conventional optimization and phase-only optimization, including a comparison to the uniform excitation case to confirm the realization of a superdirective antenna array. 
%
Finally, Section IV concludes the paper by summarizing the key findings and suggesting avenues for future research.

\begin{figure}[t] 
  \centering
  \includegraphics[height=0.42\linewidth]{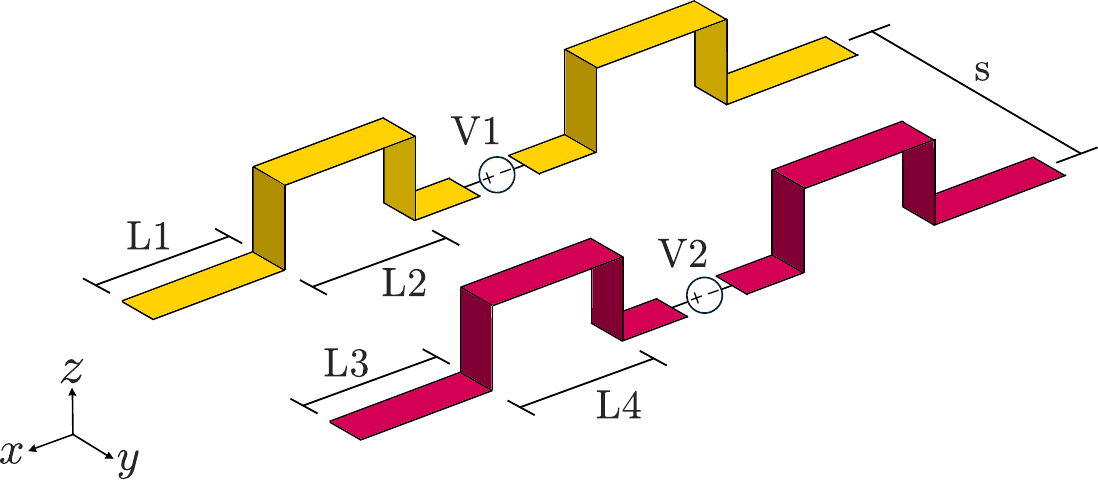}
  \caption{Two Element Meandered Dipole Antenna Array.}
  \label{Fig:1}
\end{figure}

\section{Antenna Design}

The general structure of the linear antenna array is depicted in Fig.~\ref{Fig:1} and consists of two meandered strip dipole elements that are spaced apart at a center-to-center distance, $s$, along the $y$ axis. Individually, the elements are excited in the center by the signals $V1$ and $V2$. The elements consist of lowered and raised lengths, $L1$ and $L2$ for the first element, $L3$ and $L4$ for the second element, respectively. The width of the strips is set to $6$~mm and the height of the raised lengths is set to 17~mm. 

\subsection{Single Element Analysis}

A single element is designed using the Antenna Toolbox in MATLAB to radiate at $900$ MHz on the basis of these constraints. Copper material with a conductivity of $5.8 \times 10^7$ S/m was chosen for the meandered strip. The total length of the single strip element is $125$~mm and presents a directivity of $1.99$~dBi. Compared to a half-wavelength dipole operating at the same frequency, which has a length of $167$~mm and a directivity of $2.15$~dBi. Using this simple meander, a reduction of $42$~mm or $25$\% is observed, with only a slight decrease in terms of directivity.   

\subsection{Optimization Setup}

An optimization was performed using a Genetic Algorithm (GA) implemented in MATLAB, with the goal of maximizing the realized gain of the meandered dipole array. Copper material with a conductivity of \(5.8 \times 10^7\) S/m was used for the meandered dipole elements. The GA was configured with a population size of $50$ and a maximum of $1000$ generations. The element lengths ($L_1$, $L_2$, $L_3$, $L_4$) were constrained within $20$\% of their initial values derived from the single element design (i.e., between 20 mm and 30 mm). The choice of GA as an optimization method is particularly beneficial for this type of design because of its ability to explore a wide solution space and converge toward global optima. The spacing of the elements ($s$), ranging from $0.05\lambda$ to $0.5\lambda$ (where $\lambda$ is the wavelength in free space at 900 MHz), was varied but was not part of the optimization process. Signal magnitude values for each element were bounded between 0 and 1, while the phase angle was bounded between 0° and 360°.

Three excitation schemes were evaluated: the conventional optimization to achieve superdirectivity (signal magnitude and phase angle), phase-only optimization, and uniform excitation (equal amplitude and no phase shift). Each optimization targeted a maximum realized gain in the end-fire direction ($+y$), providing insights into the impact of different excitation strategies on the performance of the two-element meandered dipole array.

The performance of the array is mathematically expressed for the two optimization schemes: conventional optimization and phase-only optimization. Additionally, the uniform excitation case is evaluated using the optimized lengths from both schemes for baseline comparison:

\vspace{0.1cm}

\subsubsection{Conventional Optimization for Superdirectivity}

\begin{equation}
\begin{aligned}
\underset{
\left\lbrace L_1, L_2, L_3, L_4, v_1, v_2 \right\rbrace 
}{\text{Maximize}}
& \quad f \left(L_1, L_2, L_3, L_4, v_1, v_2\right) \\
\text{subj. to:} 
        & \quad L_i \in [20\, \text{mm}, 30\, \text{mm}], \quad i=1,2,3,4, \\
        & \quad |v_1|, |v_2| \in  \left[0, 1\right],\\
        & \quad \phase{v_1}, \phase{v_2} \in \left[0^{\circ}, 360^{\circ}\right],
\end{aligned}
\end{equation}

\subsubsection{Phase-Only Optimization}

\begin{equation}
\begin{aligned}
\underset{
\left\lbrace L_1, L_2, L_3, L_4, \phase{v_1},\phase{v_2} \right\rbrace 
}{\text{Maximize}}
& \quad f \left(L_1, L_2, L_3, L_4, \phase{v_1}, \phase{v_2}\right) \\
\text{subj. to:} 
        & \quad L_i \in [20\, \text{mm}, 30\, \text{mm}], \quad i=1,2,3,4, \\
        & \quad \phase{v_1},\phase{v_2} \in \left[0^{\circ}, 360^{\circ}\right],
\end{aligned}
\end{equation}







\vspace{0.1cm}

Here, $f$ denotes the realized gain function, allowing us to evaluate the effect of different excitation schemes while maintaining consistent physical configurations based on the optimized element lengths.

\section{Optimization Results}

In this section, we analyze the results obtained from the conventional and phase-only optimization of the meandered dipole array. The goal is to evaluate the impact of different excitation schemes on the element lengths, signal magnitudes, and phase angles, as well as their influence on the realized gain and overall antenna performance. The conventional optimization results will be compared against those from the phase-only optimization to identify any differences in behavior, particularly in terms of the realized gain in the end-fire direction and the interaction between element spacing and excitation.

The resultant directivity ($D$) and directivity of the uniformly excited optimal geometries (Uni $D$), for the optimized realized gain ($RG$) of the conventional optimization and phase-only optimization are presented in Table~\ref{table01} and Table~\ref{table02}, respectively. Each lists the optimal element lengths ($L_1,L_2,L_3,L_4$), signal magnitudes ($|v_1|$, $|v_2|$), and phase angles ($\phase{v_1}$, $\phase{v_2}$), for various element spacings ($s$). 

\begin{figure}[t] 
  \centering
  \includegraphics[height=0.75\linewidth]{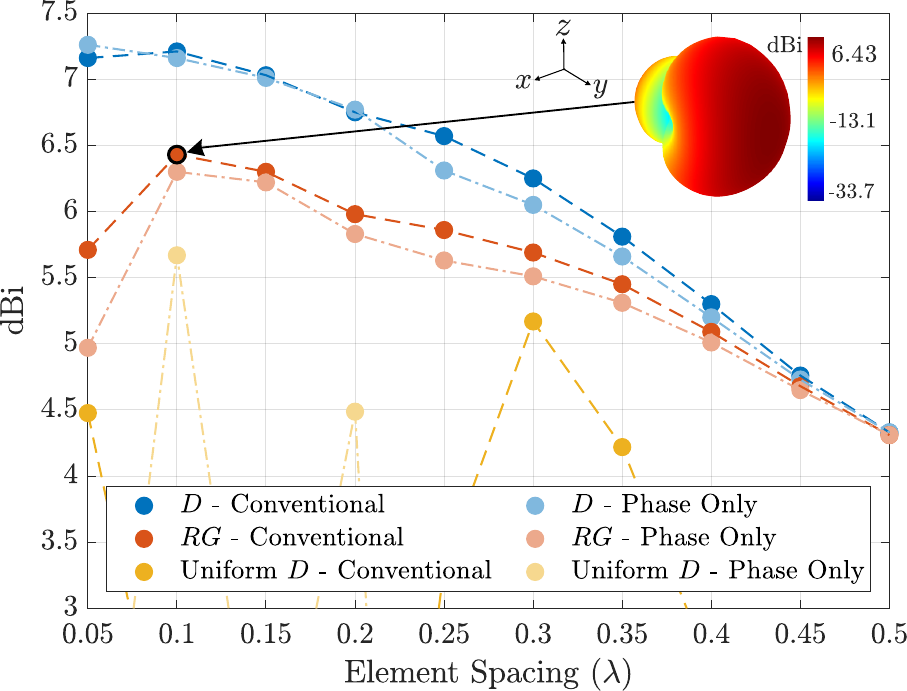}
  \caption{The simulated directivity and realized gain for the optimized (conventional and phase only) and uniform cases versus element spacing at $900$ MHz. The 3D radiation pattern for the best resultant realized gain (conventional method) when $s=0.1\lambda$ is also shown.}
  \label{Fig:2}
\end{figure}

Fig. \ref{Fig:2} illustrates the realized gain versus element spacing for the two-element meandered dipole array under both conventional optimization and phase-only optimization schemes. The results clearly highlight the performance improvements achieved through optimized excitation schemes. As observed, the realized gain reaches its peak at a spacing of 0.1$\lambda$ for both optimization strategies, with the conventional optimization showing slightly higher values compared to phase-only optimization. The uniform excitation cases, represented by the lower curves, exhibit significantly reduced gain across all element spacings, validating the importance of optimized excitation and superdirectivity according to Hansen's definition \cite{Hansen1981Fundamental}. Notably, the gradual decline in realized gain beyond the optimal spacing highlights the trade-off between element proximity and performance in compact antenna arrays. This plot underscores the effectiveness of amplitude and phase optimization in enhancing the realized gain, while phase-only optimization still delivers competitive results with fewer control parameters.%
\renewcommand{\arraystretch}{1.15} 
\vspace*{-0.5em} 
\begin{table*}[t]
\centering
\begin{tabular}{rrrrrrrrrrrr}
\toprule
 s ($\lambda$) &  $L_1$ (mm) &  $L_2$ (mm) &  $L_3$ (mm) &  $L_4$ (mm) &  $|v_1|$ &  $\phase{v_1}$ (deg.) &  $|v_2|$ &  $\phase{v_2}$ (deg.) & $D$ (dBi) & Uni $D$ (dBi) & Opt $RG$ (dBi) \\
\midrule
        0.05 &    24.6 &    23.9 &    24.7 &    20.8 &       0.5 &             248 &       0.3 &             242 & 7.16 & 4.48 & 5.71 \\
        0.10 &    27.7 &    20.3 &    26.2 &    20.5 &       0.8 &             273 &       0.6 &             282 & \textcolor{red}{7.21} & \textcolor{red}{1.35} & \textcolor{red}{6.43} \\
        0.15 &    25.0 &    23.3 &    24.8 &    23.1 &       1.0 &             108 &       0.7 &             140 & 7.03 & -8.61 & 6.30 \\
        0.20 &    24.1 &    24.1 &    24.1 &    25.1 &       1.0 &             193 &       0.4 &             256 & 6.75 & -2.03 & 5.98 \\
        0.25 &    24.2 &    24.5 &    26.6 &    20.8 &       1.0 &             351 &       0.4 &              60 & 6.57 & 3.32 & 5.86 \\
        0.30 &    26.7 &    20.7 &    25.2 &    23.5 &       0.7 &             214 &       0.3 &             344 & 6.25 & 5.17 & 5.69 \\
        0.35 &    25.2 &    22.5 &    23.0 &    26.4 &       0.9 &             244 &       0.5 &              34 & 5.81 & 4.22 & 5.45 \\
        0.40 &    23.9 &    24.8 &    21.7 &    27.4 &       1.0 &               0 &       0.6 &             149 & 5.30 & 2.54 & 5.09 \\
        0.45 &    23.6 &    24.9 &    26.0 &    21.4 &       0.3 &             360 &       0.2 &             165 & 4.76 & -2.10 & 4.68 \\
        0.50 &    27.0 &    20.4 &    23.0 &    25.6 &       0.3 &              85 &       0.3 &             263 & 4.33 & -27.14 & 4.31 \\
\bottomrule
\end{tabular}
\caption{Optimal element lengths, signal magnitudes, and phase angles for different element spacings obtained from conventional optimization, along with Directivity ($D$), Uniform Directivity (Uni $D$), and Optimized Realized Gain (Opt $RG$).}
\label{table01}
\end{table*}

\begin{table*}[t]
\centering
\begin{tabular}{rrrrrrrrrrrr}
\toprule
 s ($\lambda$) &  $L_1$ (mm) &  $L_2$ (mm) &  $L_3$ (mm) &  $L_4$ (mm) &  $\phase{v_1}$ (deg.) &  $\phase{v_2}$ (deg.) & $D$ (dBi) & Uni $D$ (dBi) & Opt $RG$ (dBi) \\
\midrule
        0.05 &    24.9 &    20.4 &    25.6 &    21.5 &       223 &       195 & 7.26 & 0.18 & 4.97 \\
        0.10 &    23.5 &    26.4 &    23.4 &    23.3 &        98 &       103 & \textcolor{red}{7.16} & \textcolor{red}{5.67} & \textcolor{red}{6.30} \\
        0.15 &    25.9 &    21.3 &    26.2 &    20.9 &       282 &       325 & 7.01 & 1.07 & 6.22 \\
        0.20 &    24.7 &    23.9 &    21.6 &    25.6 &       196 &       224 & 6.77 & 4.49 & 5.83 \\
        0.25 &    25.5 &    20.2 &    26.2 &    21.7 &       229 &       339 & 6.31 & -7.12 & 5.63 \\
        0.30 &    20.6 &    27.5 &    25.1 &    24.8 &       342 &       142 & 6.05 & -6.70 & 5.51 \\
        0.35 &    20.4 &    27.9 &    24.3 &    25.7 &        26 &       199 & 5.66 & -9.24 & 5.31 \\
        0.40 &    21.4 &    26.7 &    28.1 &    20.0 &        44 &       223 & 5.20 & -13.35 & 5.01 \\
        0.45 &    25.1 &    22.1 &    25.5 &    22.9 &       355 &       176 & 4.73 & -20.33 & 4.65 \\
        0.50 &    20.1 &    29.6 &    24.5 &    23.8 &       291 &       113 & 4.33 & -30.96 & 4.31 \\
\bottomrule
\end{tabular}
\caption{Optimal element lengths, signal magnitudes, and phase angles for different element spacings obtained from phase-only optimization, along with Directivity ($D$), Uniform Directivity (Uni $D$), and Optimized Realized Gain (Opt $RG$).}
\label{table02}
\end{table*}

\section{Conclusion}

This paper presented a comprehensive study of a compact two-element meandered dipole array aimed at achieving super-realized gain. By employing a Genetic Algorithm (GA) for optimization, two primary excitation schemes, conventional optimization and phase-only optimization were evaluated. The results showed that by carefully optimizing both the signal magnitudes and phase angles, the array could achieve a substantial improvement in realized gain, particularly in the end-fire direction. 

The phase-only optimization provided a competitive realized gain, with only minor reductions compared to the conventional optimization, suggesting that optimizing the signal phase alone can be an effective strategy in practical implementations where controlling amplitude is more challenging.

\section*{Acknowledgment}
This paper has been financed by the funding programme “MEDICUS”, of the University of Patras.

\balance

\bibliographystyle{ieeetr}
\bibliography{mybib.bib}

\end{document}